\documentclass{ptephy_v1}

\preprintnumber{XXXX-XXXX} 

\usepackage{graphicx} 
\usepackage{lineno}
\usepackage{url} 

\newcommand{\Efive}{{\rm 511\,keV}}
\newcommand{\Eone}{{\rm 1275\,keV}}
\newcommand{\Na}{{\rm {}^{22}Na}}

\begin{document}
\nolinenumbers

\title{Search for the correction term to the Fermi's golden rule in positron annihilation}


\author[1,*]{R. Ushioda}
\author[1]{O. Jinnouchi}
\author[2,3]{K. Ishikawa}
\author[4]{T. Sloan}
\affil[1]{\small Department of Physics, Faculty of Science, Tokyo Institute of Technology, Tokyo 152-8551, Japan}
\affil[2]{Department of Physics, Faculty of Science, Hokkaido University, Sapporo 060-0810, Japan}
\affil[3]{Research and Education Center for Natural Sciences, Keio University,  Kanagawa 223-8521, Japan}
\affil[4]{Department of Physics, University of Lancaster, Lancaster LA1 4YB, United Kingdom {\rm \email{ushioda@hep.phys.titech.ac.jp}}}


\begin{abstract}%
In the positron-electron annihilation process, finite deviations from the standard calculation based on the 
Fermi's Golden rule are suggested in recent theoretical work.  
This paper describes an experimental test of the predictions of this 
theoretical work by searching for events with two photons from positron annihilation 
of energy larger than the electron rest mass ($511\,{\rm keV}$). The positrons came from a 
${\rm {}^{22}Na}$ source, tagging the third photon from the spontaneous emission of ${\rm {}^{22}{Ne}^*}$
de-exitation to suppress backgrounds.
Using the collected sample of $1.06\times 10^{7}$ positron-electron 
annihilations, triple coincidence photon events in the signal enhanced energy regions are examined.
The observed number of events in two signal regions, $N^{\rm SR1}_{\rm obs}=0$ and $N^{\rm SR2}_{\rm obs}=0$ are, within a current precision, 
consistent with the expected number of events, $N^{\rm SR1}_{\rm exp}=0.86\pm0.08({\rm stat.})^{+1.85}_{-0.81}({\rm syst.})$ and 
$N^{\rm SR2}_{\rm exp}=0.37\pm 0.05({\rm stat.})^{+0.80}_{-0.29}({\rm syst.})$ from Fermi's golden rule respectively. 
Based on the $P^{(d)}$ modeling, 
the 90\% CL lower limit on the photon wave packet size is obtained. 
\end{abstract}

\subjectindex{xxxx, xxx}

\maketitle
\section{Introduction}
\textcolor{black}{In a previous publication \cite{PTEP-th2} it was shown that deviations from 
quantum electrodynamic (QED) calculations by the Fermi Golden Rule may appear in 
positron annihilation. Such deviations arise from the approximations
made in the calculations. Explicit formulae for the expected deviations
are derived fully in \cite{PTEP-th2}. In this paper an attempt to detect 
experimentally such deviations is described.}
The process of positron annihilation to two photons is a simple system and their
detection is suitable for a precision test of QED.
A transition probability of the 
quantum process,  $P(T)$, at time $T$ is formulated by the Fermi's golden rule with a certain 
approximation \cite{dirac, schiff},  i.e. $P(T)=\Gamma T$, where $\Gamma$ is the average transition rate.
A problem of this approximation has been pointed out by several theoretical considerations, and it is suggested 
that an additive constant correction term, $P^{(d)}$, is required in the formulation as in Eq.~(\ref{eq:Pd}) \cite{finite-size}.
\begin{equation}
P(T)=\Gamma T + P^{(d)}
\label{eq:Pd}
\end{equation}  
\textcolor{black}{$P^{(d)}$ term has its unusual property distinct from the golden rule, and may have been buried under
the background. Accordingly $P^{(d)}$ term has not been seriously studied in experiments. 
Therefore the golden rule has been valid when the experiments were designed for its confirmation.}
This holds true for a slow process where $T$ is large, but it is concerned to be insufficient for rapidly changing processes
where relatively sizable effect for the correction term are expected \cite{PTEP-th1}.
This was shown to be the case in positron annihilation \cite{PTEP-th2}.
Although the processes based on the correction term, $P^{(d)}$, can show a unique signature of its 
non-conserving nature of the kinetic energies, experimental confirmation of the effect has not been 
hitherto seriously pursued mainly due to its predicted broad spectrum, which would lie under the backgrounds,
preventing its manifestation. 

In the paper \cite{PTEP-th2} it is proposed that the two photon process of the position annihilation 
could manifest a sizable correction, and the feasibility of an experiment using a simple setup is discussed.   
A setup based on the ${\rm {}^{22}Na}$ radioactive source surrounded by the 
$\gamma$-ray detectors, e.g. NaI(Tl) scintillation detectors, is an ideal platform for verifying such effect. From the 
process, ${}^{22}{\rm Na} \rightarrow {}^{22}{\rm Ne}^* + e^+ + \nu,  \,e^+ + (e^-) \rightarrow \gamma_1+\gamma_2$,
two photons with the same energies in opposite directions are expected. Here, $(e^-)$ is the electrons 
resident within the materials near the source. 
\begin{figure}[!h]
\centering
\includegraphics[width=8cm]{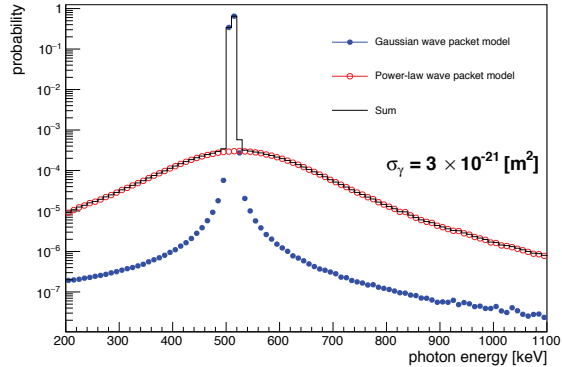}
\caption{Expected photon energy distributions based on the golden rule term plus the $P^{(d)}$ term.
\textcolor{black}{Mixture of Gaussian (blue filled points) and Power-law (red open points) wave packet models
with the ratio 99\% to 1\%, and 
the packet size of $\sigma_\gamma=3\times10^{-21}\;{\rm m^2}$, are assumed. Sum of these two models are shown (black histogram).
\textcolor{black}{If $P^{(d)}$ term is zero, $E_{\gamma_1}$ and $E_{\gamma_2}$ would be $511~{\rm keV}$ within a spread 
expected from the uncertainty principle, i.e. the natural line width. $P^{(d)}$ introduces an extra broadening of the spectrum.}
Event selection, $|E_{\gamma_1}-E_{\gamma_2}|<150~{\rm keV}$, is applied. }}
\label{fig:Pd_Effect}
\end{figure}
In Fermi's golden rule, the energies of the two photons are the electron rest mass, i.e. 
${E_{\gamma_1}=E_{\gamma_2}=511\,{\rm keV}}$, 
while with correction term, $P^{(d)}$, it can be deviate from $511\,{\rm keV}$ 
as illustrated in Fig.~\ref{fig:Pd_Effect}.

\textcolor{black}{Due to the $P^{(d)}$ term, intrinsic photon energy distribution would have long tails on
both sides around the sharp peak at the electron mass of 511 keV. 
The shape of this tail depends on the size and the 
shape of the photon wave packets. Photon wave packet size, $\sigma_\gamma$ is defined in the 
paper \cite{PTEP-th2}, and $3\times10^{-21}\;{\rm m^2}$ is assumed in Fig.~\ref{fig:Pd_Effect}.
Two photon packet shapes, Gaussian and Power-law, are assumed, and are mixed by the typical 
ratio 99 (Gaussian) to 1 (Power-law). In Fig.~\ref{fig:Pd_Effect}, these two components, together with 
their sum are separately shown. Higher energy region around 1 MeV is populated mostly by 
the Power-law model.}

In experiment, low energy range (${\rm E_{\gamma_1}=E_{\gamma_2}<511\,keV}$) suffers a huge backgrounds produced via 
Compton scattering of the $\Efive$ photons. Conversely, the high energy range  (${\rm E_{\gamma_1}=E_{\gamma_2}>511\,keV}$),
is free from such backgrounds. Exception is those from double hit pileup events discussed in Sec. \ref{sec:bg}. 
Events requiring two photons in opposite direction is named {\it 2-coincidence events}, and it 
is not enough to suppress the double hit pileup backgrounds. However, in the process of $\Na$ decays, 
the third photon (${\rm E_{\gamma_0}=1274.5\,keV}$) is emitted almost simultaneously to 
the positron annihilation via the de-excitation process of ${}^{22}{\rm Ne}^*$ nucleus. 
Tagging of this third photon is effective in significantly suppressing the double hit backgrounds.  
In this paper, the events requiring three photons, one with ${\rm 1274.5\,keV}$ and the other two 
being detected in back-to-back configuration in higher energy range, is named {\it 3-coincidence events}, and 
is treated as the signal events.  

This paper is organized as follows. In section 2, the setup of the experiment is described. 
In Section 3, the models and the methodologies of the background estimation is explained. 
Section 4, describes the data analysis. 
Section 5 summarizes the results, and the results are interpreted in Section 6. 
Section 7 concludes this new measurement.   
\section{Experimental Setup}\label{sec:setup}
\begin{figure}[!h]
\centering
\includegraphics[width=8cm]{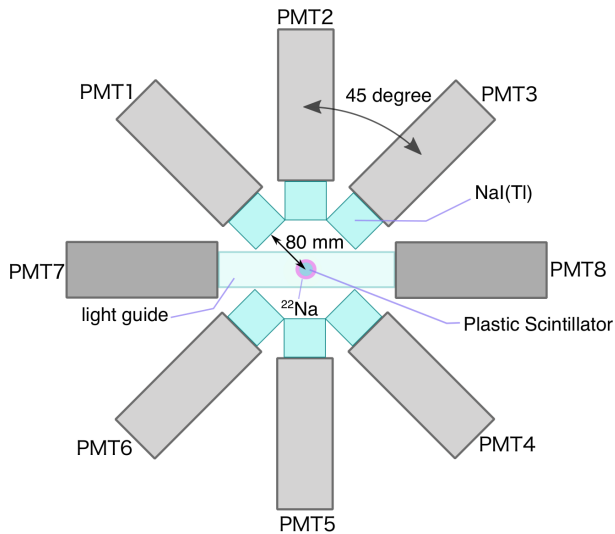}
\caption{Schematic illustration of the experimental setup top view.}
\label{fig:setup}
\end{figure}

\begin{figure}[!h]
\centering
\includegraphics[width=6cm]{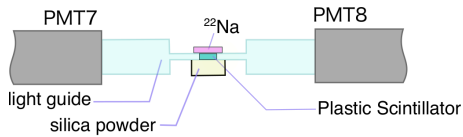}
\caption{Schematic illustration of the positron source setup side view.}
\label{fig:setup-side}
\end{figure}

Figure \ref{fig:setup} is the top view of the experimental setup. The ${}^{22}{\rm Na}$ radioactive source 
(Japan isotope center, 3 mm diameter aperture, 25 kBq) \cite{isotope}
is placed at the center, which is surrounded by the six cylindrical shaped NaI(Tl) scintillators 
(see later text for detail), named as PMT1 to 6. 
As in the figure, three pairs are made in back-to-back configuration with respect to the source, 
between them are separations of 45\,degrees each. 
Scintillators face to the center of the setup, and the distance between their 
surface and the center is 80\,mm. 
In the centre is placed a ${}^{22}{\rm Na}$ radioactive source with the positron 
tagging system as shown in Fig.~\ref{fig:setup-side}. A thin plastic scintillator plate 
(Saint-Goban, $\phi\,7.5 {\rm mm} \times t\,0.1 {\rm mm}$,  BC-408, Polybinyltoluene 1.023 g/cc \cite{plastic}) is placed beneath the  ${}^{22}{\rm Na}$ source, 
following is the small container filled with the ${\rm SiO_2}$ powder (NIPPON AEROSIL, R812 (density: $\sim 60\,{\rm mg/cm^3}$, 
Specific surface area: $230-290 {\rm m^2/g}$) \cite{aerosil}). Positrons ejected from the source scintillate in
the thin plate scintillator. The scintillation light propagates through the thin lightguides on both sides 
(CI~Industry Co. Ltd. \cite{ci}, $t$\,2\,mm$\times$30\,mm$\times$39\,mm) and the second 
lightguides ($\phi\,38\,{\rm mm} \times t\,{\rm 109.7\,mm}$), then readout by two Photomultipliers (Hamamatsu H6410). 
These PMTs are named PMT 7 and 8, and the coincidences of these are considered as the positron signals and used as 
triggers.
Positrons are trapped inside aerogel pores of the silica powder and annihilate into two photons. 
\textcolor{black}{It is assumed that the positron is captured from rest \cite{positron_at_rest1, positron_at_rest2,positron_at_rest3}. This 
gives the magnitude of the background from in-flight annihilation as less than $10^{-6}$ as 
estimated in \cite{PTEP-th2}.}
In order to reduce the contribution from positronium formation, the silica powder container 
is filled with air.  
This positron signal timing is used to suppress the accidental background by requiring 
two detections, one from the combination of the $\gamma$-ray detections and one 
from the positron signals. Additionally, requirement of positron signal in plastic scintillator can constrain the 
positron annihilation position to be inside the  ${\rm SiO_2}$ powder case. 

For the $\gamma$-ray detection, 6$\times$NaI(Tl) scintillators (OHYO KOKEN, 8B8 ($\phi 50.8\,{\rm mm} \times t\,{\rm 50.8\,mm}$) \cite{oken}) are used.
The NaI crystal is sealed in aluminum housing (${\rm 4.0\,mm}$ thick in front, ${\rm 2.8\,mm}$ thick on the sides). 
Scintillation light is readout by PMTs (Hamamatsu, H6410, H7195, H1161-50, \cite{hamamatsu}).

A CAMAC system (controller TOYO Corporation CC/NET \cite{toyo}) interfaced to a PC was used for the data acquisition.  
The analogue signals were handled with the NIM standard modules, for the digitization, timing adjustment, and taking coincidence. 
As the energy deposit of positron on the thin plastic scintillator is small, 
the outputs of PMT 7 and 8 are amplified with FastAmp (Kaizuworks Corporation, 2104\cite{kaizu}). They are digitized with low threshold, and  
timing coincidence of PMT 7 and 8 is required to mitigate the fake signals from the noise. 
One outputs from each PMT 1 to 6 are fed into discriminator module, and the signal timings are defined by the digitized 
outputs of them. A hardware coincidence of the positron signal and at least one hit out of six PMTs for the 
photons (PMT 1-6), is used as a trigger for the data taking. 
Time difference, $\delta t_i$, between positron signal and each photon detector ($i=1-6$)
is measured with the Time to Digital Converter (TDC, Technoland Corporation, C-TS103, 125 ps resolution \cite{techno}). 
Other outputs from each PMT 1 to 6 (each PMT has two anode outputs) are used to 
measure the energy depositions inside NaI(Tl).  
In order to measure the energy deposition, these outputs are subdivided into two lines, and are fed to a charge sensitive 
Analogue to Digital Converter (ADC, Hoshin Electronics Co., LTD., C009 \cite{hoshin}) with different gate widths to mitigate the pile-up events.  
Hereafter they are called ${\rm ADC_{wide}}$ (gate width ${\rm 1.2\,\mu s}$) 
and ${\rm ADC_{narrow}}$ (gate width ${\rm 100\,ns}$).
\section{Backgrounds} \label{sec:bg}
Although there is no intrinsic background source for this experiment, where two photons with large energies radiated 
in back-to-back directions, there are several potential sources which mimic the signal. 

One such background is the environmental radiations from walls of the experimental room. In order to 
reduce this, a coincidence in timing between the positron emission and the photon
detections is required. Accidental coincidence rate, $R_{acc}$, is expected to be in the form, 
$R_{acc}=R_{e^+}\times R_{\gamma}\left(h_{e^+}+{h_{\gamma}}-2h_{coin}\right)$,
where $R_{e^+}$ is the positron detection rate ($\sim 100$\,Hz), 
$R_{\gamma}$ is the environmental radiation detection rate ($\sim 0.17$\,Hz), 
$h_{e^+}$ is the pulse width of the positron signal ($\sim 100$\,ns),
 $h_{\gamma}$ is the pulse width of the photon signal ($\sim 100$\,ns),
and $h_{coin}$ is the minimum time width required for the coincidence ($\sim 10$\,ns). 
During the measured time (550 hours), 5 events are expected from the environmental background 
which concentrate in photon energies below $\Efive$. Hence the environmental radiations are 
safely ignored.  

Main background events stem from the double hits pileup, where the data acquisition is incapable to separate 
two sequential $\Na$ decays due to the finite time window of the coincidence module. These pileup events 
can be suppressed by the comparisons of the two ADC measurements for the photon energies. 
However it can only reduce the pileup event rate down to ${\it {O}}({10^{-4}})$, therefore the simulation 
based estimation is vital.

\begin{figure}[!ht]
\centering
\includegraphics[width=12cm]{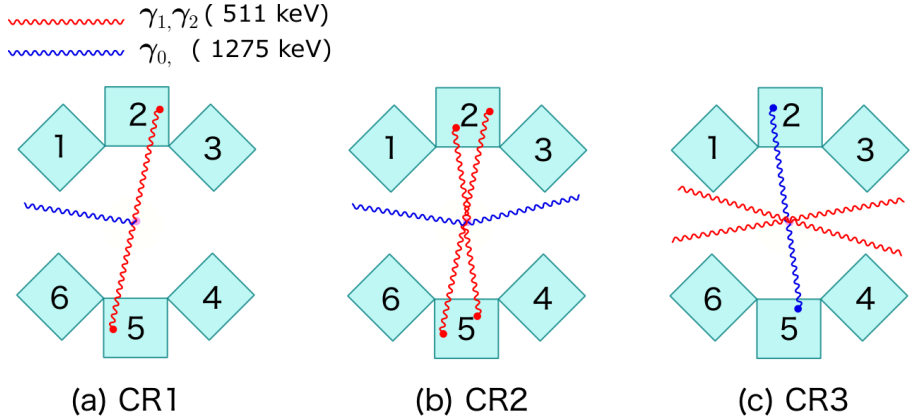}
\caption{2 coincidence event types considered in this analysis. See text for description for each drawing.}
\label{fig:event_types}
\end{figure}

Figure\,\ref{fig:event_types} shows several event configurations. 
Figure\,\ref{fig:event_types}(a) represents the nominal event case where $\Efive$ photons 
from annihilations are detected in opposite side detectors. 
Figure\,\ref{fig:event_types}(b) is the double hit background where two pairs of $\Efive$ photons are 
detected, i.e. $1022{\rm \,keV}$ energies in back-to-back detectors. 
Figure\,\ref{fig:event_types}(c) is another type of double hit background. In this case, two $\Eone$ 
photons from different decays are detected in opposite detector pair by accident. 

\begin{figure}[!ht]
\centering
\includegraphics[width=14cm, bb=0 0 749 368]{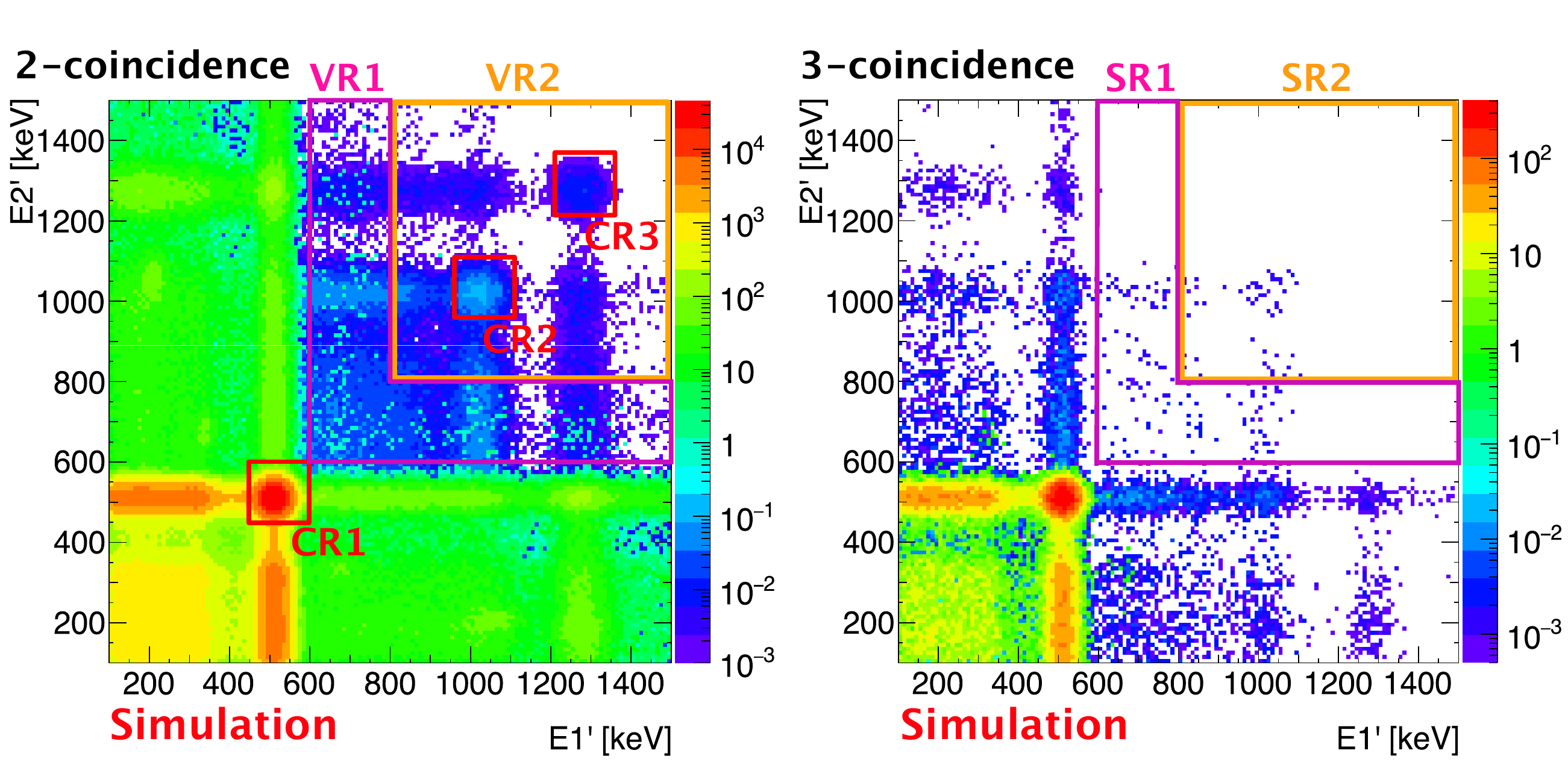}
\caption{$E'_2$ vs. $E'_1$ distributions estimated with the simulations.}
\label{fig:simulations}
\end{figure}
In order to estimate the double hit background events in 2- or 3-coincidence events, a detector 
simulation based on the Geant4 toolkit\cite{G4} has been setup. The detector components and materials in 
Fig.\,\ref{fig:setup} and Fig. \ref{fig:setup-side} are reproduced in the simulation. In single hit events, two $\Efive$ back-to-back 
annihilation photons, and a $\Eone$ photon are simulated, while in double hit events two pairs of  
back-to-back $\Efive$ photons and two $\Eone$ photons are simulated. 
In these simulations, $\Eone$ photons are emitted 
isotropically from the $\Na$ source, and the positron annihilations are generated inside the silica powder 
container.  

$2\times10^8$ simulation events are generated in both types of events. 
In order to reproduce the experimental data, single hit events and double hit events 
are summed with the proper weights.  The weights are estimated by using the 
experimental data as explained in Sec.\ref{sec:backgroundestimate}. 

In this experiment, the {\it 2-coincidence} detection is defined as follows. 
2-coincidence :  ``photons are detected in more than or equal to 2 detectors and only one pair of 
them are in opposite direction, i.e. excluding the events with two or more opposite direction pairs.''
Similarly the {\it 3-coincidence} detection is defined as follows. 
``3-coincidence : photons are detected in one opposite direction pair detectors, and in an additional
detector.  The energy deposit in additional detector, $E'_3$, should be, ${\rm 1200~keV}\le E'_3\le {\rm 1390~keV}$. 
As in the case for the 2-coincidence, the events having two or more opposite direction pairs are excluded.''
Requirement on $E'_3$ strongly suppresses the backgrounds of type (c) in Fig.~\ref{fig:event_types} 
where two {\rm 1275~keV} photons accidentally hit the opposite side detectors. 

In these coincidence events, energy depositions in the opposite direction pair are named as $E'_1$ and $E'_2$, 
where $E'_1$ is measured with one of PMT-1, 2 and 3. Similarly, $E'_2$ is the one from PMT-4,5 and 6.
In the 2-dimensional phase space of $E'_1$ and $E'_2$, three control regions (CRs) are defined. 
These CRs abound with background in 2 coincident events, and they are used 
to normalize the distributions from simulation to the measured data.
Also defined are the two signal regions (SRs), where large signal
to background ratio is expected, and two validation regions (VRs) which are used to check the validity of background 
estimation at the region close to SRs. They are defined in Table.~\ref{table:Regions}. 
\begin{table}[!t]
\caption{Definitions of the CR, SR and VR regions (see Fig. 5 for a pictorial illustration of the different named regions.). The remaining number of events out of $2\times10^8$ simulated event 
samples, i.e. single hit and double hit events, after the energy cuts, and $c_{ij}$ correction (see Sec. 4 for
definition) are shown in 4th and 5th columns. }
\label{table:Regions}
\centering
\begin{tabular}{|c|c|c|c|c|}
\hline
regions & condition & energy cut & $N_{\rm \{regions\}, single}^{\rm simul}$ &$N_{\rm \{regions\}, double}^{\rm simul}$ \\ 
\hline
CR1 & 2-coin. & 450\,keV$\le E'_1, E'_2\le$ 600\,keV &  $1.52\times10^6$ & $2.88\times10^6$  \\  
CR2 & 2-coin. & 961\,keV$\le E'_1, E'_2\le$ 1111\,keV &  $0$ & $3.50\times10^3$  \\  
CR3 & 2-coin. & 1213.5\,keV$\le E'_1, E'_2\le$ 1363.5\,keV &  $0$ & $3.02\times10^3$  \\
VR1 & 2-coin & when $E'_i\le E'_j$,  600\,keV$\le E'_i\le$ 800\,keV,  &  $2.15\times10^2$ & $2.84\times10^4$ \\
       &             & 600\,keV$\le E'_i\le$ 1500\,keV &  &  \\
VR2 & 2-coin & 800\,keV$\le E'_1, E'_2\le$ 1500\,keV & $2.1$ & $1.59\times10^4$ \\
       &             & (excluding CR2, CR3) &  &  \\
SR1 & 3-coin. & when $E'_i\le E'_j$,  600\,keV$\le E'_i\le$ 800\,keV,  & $0$ &  $2.40\times10^2$ \\ 
       &             & 600\,keV$\le E'_i\le$ 1500\,keV &  &  \\
SR2 & 3-coin. & 800\,keV$\le E'_1, E'_2\le$ 1500\,keV  &  $0$ & $1.07\times10^2$ \\
\hline
\end{tabular}
\end{table}
The definition of these regions can be graphically confirmed in Fig.\,\ref{fig:simulations}.
CR1 is dominated by those events in which both $\Efive$ photons are photoelectrically 
absorbed (Fig.\,\ref{fig:event_types}(a)). The number of events in CR1, $N_{\rm CR1}^{\rm data}$, is used to normalize the simulation 
to the observed data. CR2 corresponds to the double hit events in which two pairs of $\Efive$
photons are all photoelectrically absorbed (Fig.\,\ref{fig:event_types}(b)). 
Number of events in CR2,  $N_{\rm CR2}^{\rm data}$, is used to estimate the amount of pileup.
On the other hand, CR3 corresponds to the double hit events in which two $\Eone$ photons from 
different events are photoelectrically absorbed (Fig.\,\ref{fig:event_types}(c)). 
Number of events in CR3,  $N_{\rm CR3}^{\rm data}$, is used to estimate 
the amount of accidental background. 
VRs suffer from less backgrounds, but they still contain good amount of backgrounds. These 
are mainly from those events where one, or both of two photons in CR2 or CR3 are Compton scattered. 
These regions can potentially contain contamination from the signal events but are expected to be 
largely dominated by the background, thus it is considered to be safe to use these as  VRs. 
SR1 and SR2 are the same as VR1 and VR2 respectively except they require 3-coincidence, 
and backgrounds are largely suppressed. 
\section{Data Analysis} \label{sec:backgroundestimate}
$1.06\times10^7$ events were recorded in 550 hours, which corresponds to the average data acquisition rate of $5.4$ Hz.
Data were taken in 13 runs. 

The energy scale of the detectors were calibrated in situ using the photo-electric absorption  
peaks ($\Efive, \Eone$) from the $\Na$ source for each run. A linear function is used for the calibration. 
The TDC is a clock counter type module, and the linearity of $\pm500~{\rm ps}$
is guaranteed. Hence no calibration runs were taken for TDC. 

In order to retain adequate quality of data, the following preselections are applied. 
Since the signal is prompt positron annihilation, the time difference between the trigger (positron annihilation) 
and the photon detection, is constant. $|\delta t|<$1\,nsec around the mean timing is required.     
The slewing effect is confirmed to be very small for relevant energy range ( $>$ 200\,keV), hence the slewing correction 
for the photon detection, i.e. energy dependent timing correction, is not applied in this analysis.

The standard deviation of the time resolution in each photon detector is 1.2\,nsec, and the time difference between photon detections within $\pm$6\,nsec are used as the coincidence events. 
As described in Sec.~\ref{sec:setup}, for each photon detection, two ADC measurements with different gate widths 
are utilized to suppress the pileup events. For the signal regions ($E^{(i)}>600\,{\rm keV}$, $i$=1-6), a cut on event 
$E^{(i)}_{\rm wide}-E^{(i)}_{\rm narrow}<100\,{\rm keV}$ is applied. 
After the preselections, the event selections for control and validation regions are applied. 

Table~\ref{table:CRresults} summarizes the number of remaining events after the event selection for 
control regions as defined in Table~\ref{table:Regions}. 
Mis-modeling of the simulation is evaluated by the comparison of data and simulated distributions 
in the region next to CR1 (450\,keV$\le E'_i\le$ 600\,keV, 600\,keV$\le E'_j\le$ 1500\,keV, $i,j=1,2, i\ne j$).
The correction factors are extracted from this comparison, 
and are applied to the simulated events to compensate the differences. 
The regions ($E'_1, E'_2 > 600\,{\rm keV}$) are binned in 2D-matrix in every 100\,keV, 
and each region is corrected by the correction factor, $c_{ij}=c^{(1)}_i\times c^{(2)}_j$, where
$c^{(1)}_i, c^{(2)}_j$ turned out to be values between 0.7 to 1.3 depending on the energy. 
$c^{(1)}_i, c^{(2)}_j (i, j = 1, 2, 3, \cdots)$ are the factors for $E'_1, E'_2$ respectively.

Using these experimentally obtained numbers, the accidental background rate and the 
pileup event rate are evaluated as follows. 
The accidental backgrounds originate from the double hit events
and appear in CR3 where two $\Eone$ photons are accidentally detected in opposite side detectors in the time window
of this experiment. 
The accidental background event rate i.e. the ratio of double hit events to single hit events, 
named $\alpha$, is determined from the equation,
$N_{\rm CR3}^{\rm data}/N_{\rm CR1}^{\rm data}=(N_{\rm CR3,single}^{\rm simul}+\alpha N_{\rm CR3,double}^{\rm simul})/(N_{\rm CR1,single}^{\rm simul}+\alpha N_{\rm CR1,double}^{\rm simul})$. $\alpha$ is 
estimated to be $\alpha=(3.35^{+4.43}_{-2.16})\times10^{-4}$. The estimated number of events in CR1, CR2 and CR3 after taking $\alpha$
into account, are $N_{\rm CR1,\alpha}^{\rm estim}=1.52\times10^6, N_{\rm CR2,\alpha}^{\rm estim}=1.17$, and $N_{\rm CR3,\alpha}^{\rm estim}=1.01$ respectively. 
Since the rejection power for pileup backgrounds after the timing coincidence 
is weaker than for accidental background, more pileup events are expected in CR2 and in 
associated lower energy range. 
In order to estimate this amount, another type of simulation sample, called {\it semi-double 
event}, containing two pairs of $\Efive$ photons and one $\Eone$ photon per event, are generated, aiming to 
simulate the pileup events which appear in CR2 and not in CR3. 
The remaining number of events in 
CR1 and CR2, out of generated $2\times10^8$ semi-double events, 
$N_{\rm CR1,semi}^{\rm simul}$, $N_{\rm CR2,semi}^{\rm simul}$, are 
found to be, $N_{\rm CR1,semi}^{\rm simul}=2.92\times10^6$, $N_{\rm CR2,semi}^{\rm simul}=2.85\times10^3$. 
The ratio of pileup events to accidental events in CR2, $\beta$, is determined from the equation, 
$N_{\rm CR2}^{\rm data}/N_{\rm CR1}^{\rm data}=(N_{\rm CR2,\alpha}^{\rm estim}+\beta N_{\rm CR2,semi}^{\rm simul})/(N_{\rm CR1,\alpha}^{\rm estim}+\beta N_{\rm CR1,semi}^{\rm simul})$, and found to be $\beta=(2.08^{+2.57}_{-1.93})\times10^{-3}$.
\begin{table}[!t]
\caption{The number of events in 2-coincidence control regions after the selections. }
\label{table:CRresults}
\centering
\begin{tabular}{|c|c|c|}
\hline
regions & name & number of events \\ 
\hline
CR1 &   $N_{\rm CR1}^{\rm data}$ & $3.01\times10^6$ \\  
CR2 &   $N_{\rm CR2}^{\rm data}$ & $14$ \\  
CR3 &   $N_{\rm CR3}^{\rm data}$ & $2$ \\
\hline
\end{tabular}
\end{table} 
\begin{figure}[!ht]
\centering
\includegraphics[width=14cm, bb=0 0 745 368]{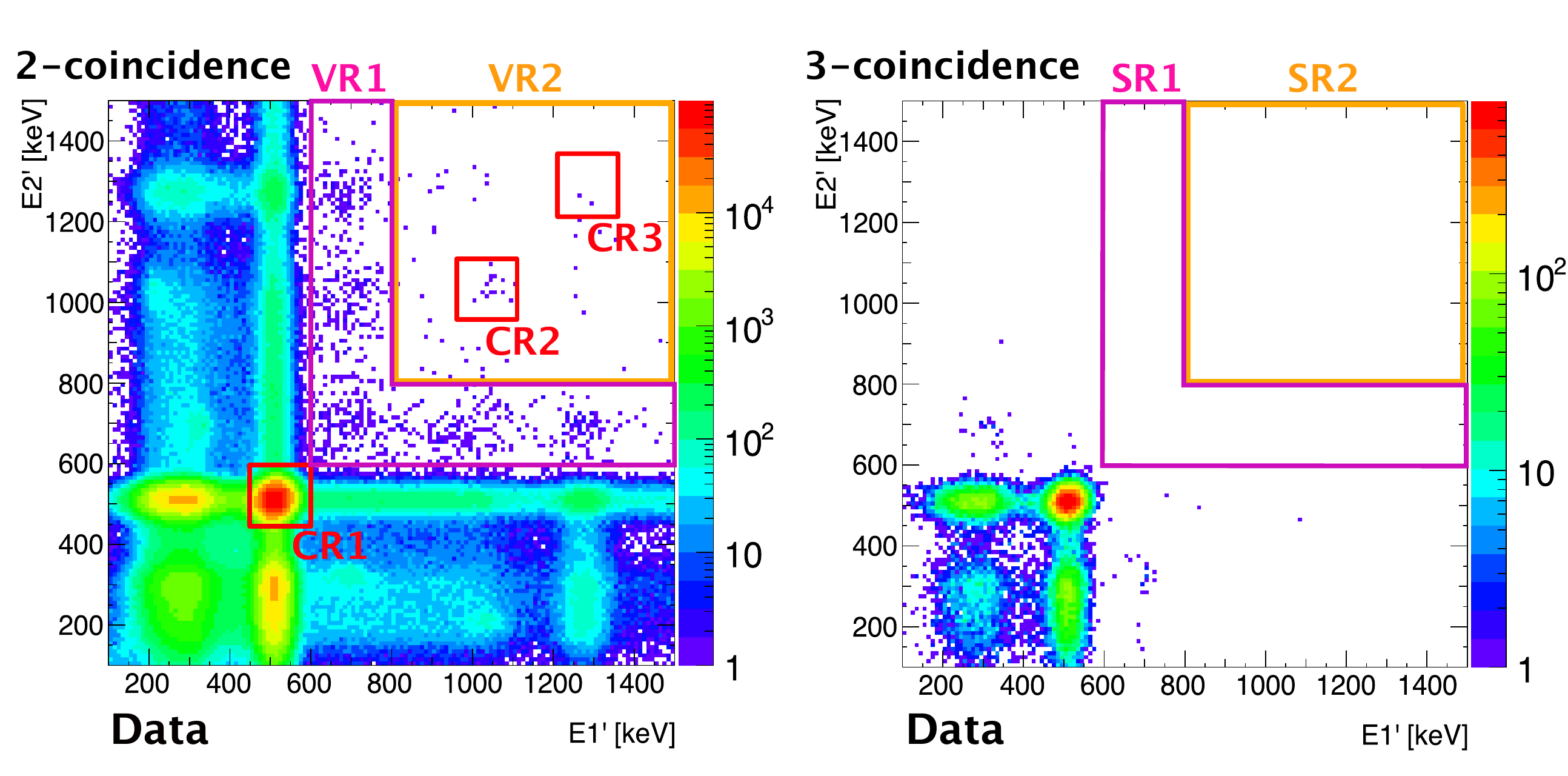}
\caption{$E'_2$ vs. $E'_1$ distributions obtained with data.}
\label{fig:data}
\end{figure}
\begin{table}[!t]
\caption{The normalized simulated number of events in various regions. 
The subscripts '$\alpha\beta$' represents the corrections made for parameters, $\alpha$ and $\beta$.}
\label{table:Estimates}
\centering
\begin{tabular}{|c|c|}
\hline
regions & $N_{\rm \{regions\}, \alpha\beta}^{\rm simul}$  \\ 
\hline
CR1 &  $1.53\times 10^6$    \\  
VR1 & $2.56\times10^2$  \\
VR2 & $1.53\times10^1$ \\
SR1 & $2.79\times10^{-1}$ \\
SR2 & $1.21\times10^{-1}$ \\
\hline
\end{tabular}
\end{table} 
Based on the parameters, $\alpha$ and $\beta$, the numbers of simulated events in various 
regions are normalized. Obtained values are summarized in Table~\ref{table:Estimates}.

The detector components are aligned, and the precision of the geometrical alignments are expected to be within $\pm 1$\,mm. 
The systematics errors related to following items are estimated. 
The distance between the photon detectors and the 
center of the setup, the tilt of the setup, i.e. the angular deviation of the detectors from the setup design,  
the actual location of the annihilations inside the powder case, etc. All these are checked using the simulation, 
and confirmed to have negligible contributions to the result. 
\section{Results}
VR1 is used to check the validity of the background estimations. In VR1, number of events in data
is found to be consistently higher than the estimation (mean factor $1.56$), hence the constant correction  
is applied on the estimation (VR1 correction) to remove this difference. A remained bin by bin fluctuation is approximately 
40\%, which is counted as the systematic errors.  
After the VR1 correction, VR2 is used for the final validity check of the method. $46$ events are observed in data ($N_{\rm VR1}^{\rm data}$), 
where $N_{\rm VR1}^{\rm estim, corre}=$ $47.0^{+8.6}_{-4.5}({\rm stat.})^{+90.9}_{-27.5}{(\rm syst.)}$ events are expected from the background estimation. A superscript 'corre' represents the corrections made for parameters $\alpha, \beta$, and the factor 1.56.

The signal regions are unblinded after this confirmation. 
Figure~\ref{fig:data} is the observed $E'_2$ vs. $E'_1$ distributions of the 2-coincidence (left) and 3-coincidence (right) events. 
Table~\ref{table:SRresults} summarizes the results in two SRs. 
The expected values in two SRs are obtained as, $N_{\rm SR{\it i}}^{\rm exp, corre}=N_{\rm SR{\it i}, \alpha\beta}^{\rm simul}\times (N_{\rm CR1}^{\rm data}/N_{\rm CR1,\alpha\beta}^{\rm simul})\times1.56\,\,\,(i=1,2)$.

0 events are observed in both signal regions, where $N_{\rm SR1}^{\rm exp, corre}=$$0.86\pm0.08({\rm stat.})^{+1.85}_{-0.81}({\rm syst.})$
and $N_{\rm SR2}^{\rm exp, corre}=$$0.37\pm0.05({\rm stat.})^{+0.80}_{-0.29}({\rm syst.})$ background events are predicted in SR1 and SR2, respectively.
With the current experimental precision, the results are consistent with the background prediction expected 
from the Fermi's golden rule.   Relatively large systematic errors stem from the propagated uncertainties of the $\alpha, \beta$ 
parameters, which originated from the limited statistics in  $N_{\rm CR2}^{\rm data}$ and $N_{\rm CR3}^{\rm data}$.

\begin{table}[!t]
\caption{The number of events in signals regions. The second column is the observed number of events in data, and 
the third column is the expected number of events from background estimation.}
\label{table:SRresults}
\centering
\begin{tabular}{|c|c|c|}
\hline
regions & $N_{\rm SR{\it i}}^{\rm data}(i=1,2)$ & $N_{\rm SR{\it i}}^{\rm exp, corre}(i=1,2)$  \\ 
\hline
SR1 &  $0$ & $0.86\pm0.08({\rm stat.})^{+1.85}_{-0.81}({\rm syst.})$  \\  
SR2 &  $0$ & $0.37\pm0.05({\rm stat.})^{+0.80}_{-0.29}({\rm syst.})$ \\  
\hline
\end{tabular}
\end{table} 
\section{Interpretation}
In the article \cite{PTEP-th2}, the relative size of $P^{(d)}$ correction term for the positron annihilation is 
predicted as a function of the photon wave packet size, $\sigma_\gamma$, for 
power-law and Gaussian wave functions. Figure 1 of the article \cite{PTEP-th2}, illustrates 
the expected ratio of events from $P^{(d)}$ per positron in SR2. Since 0 events were observed in SR2, 
90\% CL upper limit, $N^{\rm SR2, 90CL}_{\rm obs}$, is 2.30 events \cite{PDG:statistics}.   
The efficiencies of detecting photons in the NaI(Tl) scintillator with $800~{\rm keV}$ threshold, $\varepsilon_{\gamma}$, 
are estimated with simulation. It is gradually increasing as a function of the incident photon energy, e.g. 
for the photon energies of $1\,{\rm MeV}$ and $1.5\,{\rm MeV}$, they are estimated to be 28\% and 38\% respectively.
Upper limit of the $P^{(d)}$ per positron in SR2 is, 
$P^{(d)}_{\rm SR2, 90CL}=(N^{\rm SR2, 90CL}_{\rm obs}-0.37)/N^{\rm 3\mathchar`-coincidence}_{\rm 511keV}/\varepsilon_{\gamma}^2$,
where $N^{\rm 3\mathchar`-coincidence}_{\rm 511keV}$ is the number of events of the positron annihilation in 3 coincidence events observed around the photo electric peak in both $E'_1$ and  $E'_2$, and is observed to be 23085. 
Using the mean value, $\varepsilon_{\gamma}=33\pm5\%$, 
$P^{(d)}_{\rm SR2, 90CL}=(7.68\pm0.20({\rm stat.})^{+3.94}_{-2.67}({\rm syst.}))\times 10^{-4}$ is obtained, 
whose central value corresponds to 
\textcolor{black}{the lower limit of $\sigma_\gamma=1.0\times 10^{-21} {\rm m^2}$
with 99 to 1 composition ratio of Gaussian and Power-law wave function models  \cite{PTEP-th2}.
}
%
\section{Conclusion}
In investigating the correction term of the Fermi's golden rule, the experimental test 
to search for the positron annihilation events with high energy two photons is carried out. 
0 events are observed in two signal regions, where 
$0.86\pm0.08({\rm stat.})^{+1.85}_{-0.81}({\rm syst.})$,  
$0.37\pm0.05({\rm stat.})^{+0.80}_{-0.29}({\rm syst.})$ are expected respectively. The result in the second 
signal region is interpreted with the $P^{(d)}$ model, yielding 90\% CL 
\textcolor{black}{lower limit on the photon wave packet size of 
$\sigma_\gamma=1.0\times 10^{-21} {\rm m^2}$.}
This is the first experiment to look into this correction term using the positron annihilation.

\section*{Acknowledgment}
The authors thank A. Kubota, T. Matsuzaki for useful discussions.

\begin{thebibliography}{9}
\bibitem{PTEP-th2}
K. Ishikawa, O. Jinnouchi, A. Kubota, T. Sloan, T.H. Tatsuishi, and R. Ushioda, Prog. Theor. Exp. Phys. {\bf 2019}, no. 3, 033B02, (2019).
\bibitem{dirac} P.A.M. Dirac, Pro. R. Soc. Lond. A {\bf 114}, 243, (1927).
\bibitem{schiff} L.I. Schilff, Quantum Mechanics (McGRAW-Hill book, company, Inc. New York), (1955).
\bibitem{finite-size} K. Ishikawa and Y. Tobita, Prog. Theor. Exp. Phys. {\bf 2013}, no. 7, 073B02, (2013).
\bibitem{PTEP-th1} 
K. Ishikawa and K. Oda, Prog. Theor. Exp. Phys. {\bf 2018} no. 12, 123B01, (2018).
\bibitem{isotope} Japan Radioisotope Association \url{https://www.jrias.or.jp/e/index.html}
\bibitem{plastic} Saint-Goban Crystals, \url{https://www.crystals.saint-gobain.com}
\bibitem{aerosil} NIPPON AEROSIL CO., LTD.,  \url{https://www.aerosil.com/product/aerosil/en}
\bibitem{ci} Japan CI~Industry Co. Ltd., \url{http://www.cikogyo.co.jp}
\bibitem{positron_at_rest1} S. Tanuma, C. J. Powell, and D. R. Penn, J. Appl. Phys. {\bf 103}, 063707 (2008).
\bibitem{positron_at_rest2} J {\v{C}}{\'{\i}}{\v{z}}ek et al., Journal of Physics: Conference Series {\bf 505}(2014), 012043
\bibitem{positron_at_rest3} H. A. Bethe and R. H. Fowler, Proc. Roy. Soc. London, Ser. A: Math. Phys. Sci. {\bf 150}, 129 (1935).
\bibitem{oken} OHYO KOKEN KOGYO CO., LTD., \url{http://www.oken.co.jp/en/index.html}
\bibitem{hamamatsu} Hamamatsu Photonics K. K., \url{https://www.hamamatsu.com/jp/en/index.html}
\bibitem{toyo} TOYO Corporation, \url{https://www.toyo.co.jp/english/}  
\bibitem{kaizu} Kaizuworks Corporation, \url{http://www.kaizuworks.co.jp/index-English.html}
\bibitem{techno} Techno Corporation, \url{http://www.tcnland.co.jp/?lang=en}
\bibitem{hoshin} Hoshin Electronics Co., LTD., \url{https://www.kagaku.com/hoshin/english.html}
\bibitem{G4} Geant4 Collaboration, Nucl. Instrum. Meth. A {\bf 506}, 250, (2003).; 
Geant4 Collaboration, IEEE Trans. Nucl. Sci. {\bf 53} No. 1, 270, (2006).; Geant4 Collaboration, Nucl. Instrum. Meth. A {\bf 835}, 186, (2016).
\bibitem{PDG:statistics} M. Tanabashi et al. (Particle Data Group), Phys. Rev. D {\bf 98}, 030001 (2018).


\end{thebibliography}

\end{document}